\documentclass[conference]{IEEEtran}
\usepackage{graphicx}
\usepackage{amsfonts}
\usepackage{amssymb}
\usepackage[english]{babel}
\usepackage{amsmath,amsfonts,amssymb}
\usepackage{verbatim}
\usepackage{amsbsy}
\usepackage{algorithm}
\usepackage{algorithmic}
\usepackage[T1]{fontenc}		
\usepackage[utf8]{inputenc}		
\usepackage{hyperref} 
\usepackage{placeins} 
\usepackage{makeidx}
\usepackage[numbers,sort&compress]{natbib}
\begin{document}

\title{Fiber-based biphoton source with ultrabroad frequency tunability}

\author{\IEEEauthorblockN{$\textrm{Santiago Lopez-Huidobro}^{1,2,^*}$, $\textrm{Markus Lippl}^{1,2}$, $\textrm{Nicolas Y. Joly}^{1,2,3}$, $\textrm{Maria V. Chekhova}^{1,2}$}
\IEEEauthorblockA{$^1$ Max Planck Institute for the Science of Light, Staudtstraße 2, 91058 Erlangen, Germany}{$^2$ University of Erlangen-Nuremberg, Staudtstraße 7/B2, 91058 Erlangen, Germany}
\IEEEauthorblockA{$^3$ Interdisciplinary Centre for Nanostructured Films, Cauerstraße 3, Erlangen, Germany\\
$^*$ santiago.lopez-huidobro@mpl.mpg.de}
}


\maketitle

\thispagestyle{plain}
\pagestyle{plain}

\begin{abstract}
Tunable biphotons are highly important for a wide range of quantum applications. For some applications, especially interesting are cases where two photons of a pair are far apart in frequency. Here, we report a tunable biphoton source based on a xenon-filled hollow-core photonic crystal fiber. Tunability is achieved by adjusting the pressure of the gas inside the fiber. This allows us to tailor the dispersion landscape of the fiber, overcoming the principal limitations of solid-core fiber-based biphoton sources. We report a maximum tunability of $120\,$THz for a pressure range of $4$ bar with a continuous shift of $30\,$THz/bar. At $21\,$bar, the photons of a pair are separated by more than one octave. Despite the large separation, both photons have large bandwidths. At $17\,$bar, they form a very broad ($110\,$THz) band around the frequency of the pump.

\end{abstract}

\IEEEpeerreviewmaketitle
\bigskip
The study of nonclassical states of light is an important part of the quantum revolution of the $20^{\textrm{th}}$ century \cite{clark2021special}. Over the last years, one of the most demanded  states is entangled photon pairs (biphotons) useful for fundamental studies and quantum optics applications like quantum communication and metrology~\cite{gisin2007quantum,giovannetti2011advances}. Special interest has grown towards quantum sensing, quantum-optical coherence tomography \cite{lemos2014quantum,abouraddy2002quantum}, and spectroscopy with undetected photons based on the induced coherence effect~\cite{zou1991induced,kalashnikov2016infrared,novikova2020study}. The latter enables one to probe an object at a certain wavelength by detecting not the photon interacting with the object, but the photon entangled to it, at a different wavelength. Thus, no detectors are needed for the probe photons, giving a significant technological advantage when the spectral range of the probe photons is far from the detectors' optimal efficiency. These methods require tunable biphotons with an adjustable and potentially very large spectral separation. Additionally, applications like optical coherence tomography (OCT) --- besides the large spectral separation between the biphotons --- require a large bandwidth  because the axial resolution is inversely proportional to the spectral width~\cite{vanselow2019ultra,valles2018optical}. Achieving this combination of large bandwidth, broad frequency separation, and tunability is a challenge.

Biphotons are commonly obtained through spontaneous parametric-down conversion (SPDC) in non-centrosymmetric crystals or via four-wave mixing (FWM) in materials with $\chi^{(3)}$ nonlinearity. In the last few years, solid-core optical fibers became a promising platform for generating biphotons using FWM. Some of their advantages are  large nonlinear interaction lengths~\cite{wang2001generation} and easy integrability into optical networks. However, the frequency tuning of these sources requires challenging and hard-to-access methods, such as heating or stretching the fiber \cite{ortiz2017spectral} or tuning the pump wavelength \cite{pearce2020heralded}. On top of that, the performance of these sources is often degraded by their intrinsic Raman-scattering noise.  Hollow-core photonic crystal fibers (HC-PCF) filled with noble gases are an excellent source of tunable biphotons: their dispersion landscape and, in particular, their zero-dispersion wavelength (ZDW) can be tuned by changing the gas pressure and they are free from Raman noise~\cite{nold2010pressure,finger2015raman,cordier2020raman}. 

The ZDW tunability in HC-PCFs is especially important as it allows to combine the large bandwidth of photons with their large spectral separation, within a range of wavelengths.  

Typically, the greater the frequency separation between the photons of a pair, the larger their group-velocity mismatch and the smaller the bandwidth of each photon. As a remedy, Vanselow \textit{et} \textit{al.} proposed to group-velocity match the photons in a pair by generating them on both sides of the ZDW, and demonstrated a crystal-based source with an ultra-large bandwidth of $25\,$THz~\cite{vanselow2019ultra}. However, the ZDW of a crystal cannot be tuned, thus restricting the applicability of this method. Gas-filled HC-PCFs overcome this restriction.

In FWM,  the signal ($s$) and idler ($i$) photons with frequencies $\omega_{s,i}$ are symmetrically  generated around the frequency of the pump beam ($\omega_p$). These frequencies are determined by the energy conservation and the nonlinear phase-matching conditions as
\begin{subequations}
\begin{align}
2\omega_p  &= \omega_s +\omega_i 
\label{FWMenergy},\\
2\beta_p  &= \beta_s + \beta_i +2\gamma P_P .  \label{FWMmomentum}
\end{align} \label{FWMeYm}
\end{subequations}
Here, $\beta_{p,s,i}$ are the propagation constants of the pump, signal, and idler, respectively. For the case of a gas-filled HC-PCF, they can  be expressed as $\beta_{p,s,i} =  k_0(\omega_{p,s,i}) n_{\text{eff}}(\omega_{p,s,i},T,p)$. The first factor is the wavevector in vacuum and the second is the effective refractive index,
which is also a function of the temperature ($T$) and pressure ($p$) of the noble gas \cite{travers2011ultrafast, nold2010pressure}. $P_P$ is the peak power of the pump and $\gamma$ is the nonlinear coefficient of the fiber. The generated biphoton state at the output of the PCF can be written as~\cite{shih2020introduction}
\begin{equation}
|\psi \rangle = \iint d\omega_s d\omega_i\,  F(\omega_s, \omega_i) |\omega_s, \omega_i\rangle,
\end{equation}
where, the term $ |\omega_s, \omega_i \rangle$ describes a photon pair. The function $F(\omega_s,\omega_i)$ is the joint spectral amplitude (JSA), and it corresponds to the complex probability amplitude to generate a photon pair with one photon at frequency $\omega_s$ and the other photon at frequency $\omega_i$.

In this letter, we experimentally demonstrate a biphoton source with an ultrabroad frequency tunability based on FWM in a gas-filled HC-PCF. Adjusting the pressure inside the PCF allows us to fully control the process in a simple and efficient way. For the generated biphotons, we study the joint spectral intensity (JSI), $|F(\omega_s,\omega_i)|^2$, and the single-photon spectra by means of single-photon fiber spectroscopy \cite{valencia2002entangled,avenhaus2009fiber}.
 
\begin{figure}[h!]
\centering
\includegraphics[width=0.95\linewidth]{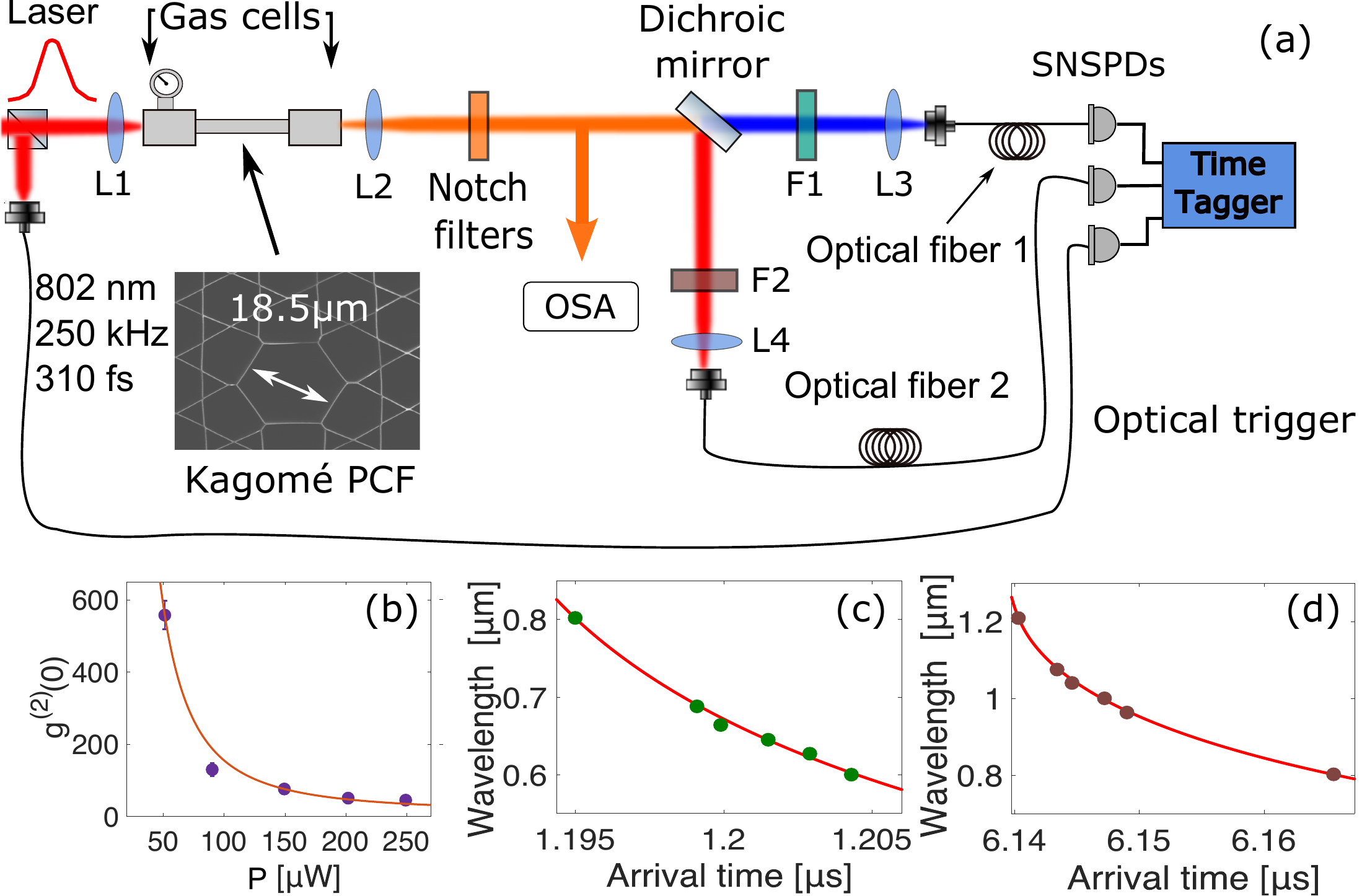}
\caption{(a)  Tunable biphoton source and the single-photon fiber spectroscopy setup. The pump beam is launched into the kagomé PCF with lens L1. The output coupling is done through lens L2, and immediately after, the notch filters suppress the pump radiation.  The dichroic mirror splits the signal and idler photons, which are respectively coupled into optical fibers 1 and 2 using lenses L3 and L4. The radiation between 750~nm and 850~nm, including the pump, is suppressed with a sequence of a short-pass (F1) and a long-pass (F2) filters. Photon pairs are detected with SNSPDs. Inset: Scanning electron micrograph of the used kagomé PCF. (b) Pump-power dependence of the second-order correlation function at a constant pressure of 19.7 bar. (c) and (d) Calibration curves using the reference spectra for the signal and idler beams measured with an optical spectrum analyzer (OSA).}
\label{fig:setup}
\end{figure}

Our biphoton source is based on a $30\,\mathrm{cm}$ long kagomé PCF  with a core diameter of $18.5 \: \mu$m (flat-to-flat) and a core wall thickness of $240\,\mathrm{nm}$. Both ends of the HC-PCF are mounted into gas cells allowing the in-coupling of the light as well as the filling of the fibre with xenon gas. A scanning electron micrograph of the used PCF is shown in the inset of Fig. \ref{fig:setup}(a). The fiber offers a very broad (several hundred $\mathrm{nm}$) transmission window at moderate losses (few dB/m) \cite{finger2015raman}, along with small anomalous group velocity dispersion (GVD), which is compensated with the normal dispersion of the xenon gas. As a result, the ZDW depends on the pressure of xenon leading to an ultrabroad frequency tunability of the signal and idler photons at high pressures \cite{azhar2013nonlinear}. A pump beam (Ti:Sapphire laser), with a central wavelength of $802\:\textrm{nm}$, a pulse duration of $310\: \textrm{fs}$, and a repetition rate $\text{R}_p = 250\,$kHz, is launched into the kagomé PCF with an aspheric lens (L1) with a focal length $\text{f}=50\,$mm. The pump power is controlled using a half-wave plate and a polarizing beamsplitter (not shown). For the output coupling, we use an $\text{f} = 40\,$mm aspheric lens (L2). Consecutively, the pump is filtered out with four notch filters --- the first two centered at $785\,$nm  and the second two at $808\,$nm. The bandwidth of each filter is $ 34\,$nm. The signal and idler beams are split with a dichroic mirror (cut-on wavelength $805\:\textrm{nm}$) and coupled into $243\,$m (Optical fiber 1) and $1254\,$m (Optical fiber 2) of SMF-28 commercial fibers using $\text{f}=35\,$mm aspheric lenses (L3 and L4). Spurious noise from the pump is further suppressed using short- (F1) and long-pass (F2) filters (Fig. \ref{fig:setup}(a)). The biphotons are eventually detected with  infrared superconducting nanowire single-photon detectors (SNSPD), whose output pulses are sent to a Time Tagger to register the single-count rates ($N_s$, $N_i$) of the signal and idler photons, as well as the rate of coincidence detection events ($N_{si}$) between the two. For all the experiments, the detection is performed in a time window of $100\,\mathrm{ps}$.

We characterized the performance of the source by measuring the normalized second-order correlation function ($g^{(2)}$) at different pump powers and a fixed pressure of $19.7\,\mathrm{bar}$.  The $g^{(2)}$ function can be written as \cite{ivanova2006multiphoton,spring2013chip}
   \begin{equation}\label{normg2}
   g^{(2)}(\tau) = \frac{N_{si}(\tau)\text{R}_p}{N_s N_i},
\end{equation}
where $N_{si}(\tau)$ is the rate of two-photon detection events separated in time by $\tau = t_1- t_2$. For the normalized second-order correlation function measurement, we only focused on the case of $\tau = 0$. Fig. \ref{fig:setup}(b) shows the obtained $g^{(2)}(0)$ values as a function of the power. We obtained a maximum value of $\sim 600$ and a decrease with the average pump power $P$, $g^{(2)}(0) \propto 1 + 1/P^2$ (fitted solid-red line). This behaviour is a fingerprint of biphoton generation \cite{ivanova2006multiphoton}. The large values of $g^{(2)}(0)$ demonstrate the absence of spurious noise, including the spontaneous Raman scattering, overcoming one of the main limitations of biphoton sources based on solid-core fibers. Moreover, the value $g^{(2)}(0) -1$, also called coincidence-to-accidentals ratio, is a measure of the source quality, and  the obtained  $g^{(2)}(0) \gg 1$ values show the potential for quantum technologies applications~\cite{lugani2020spectrally}.

The biphoton source was tuned by adjusting the pressure of the xenon filling the HC-PCF from $17.0$ up to $21.0\,\mathrm{bar}$ in steps of $0.5\,\mathrm{bar}$ and with a constant average pump power of $0.2\,\mathrm{mW}$. Over this whole range of pressure, the system exhibits normal dispersion at the pump wavelength $\lambda=802\, $nm. At these conditions the input pump yields the generation of sidebands through FWM. Operating the system in the FWM regime is  crucial for the biphoton generation since  the wavelengths of the signal and idler photons can be broadly separated. Moreover, the further the pump from the ZDW, the less effect of the pump power on the signal and idler wavelengths. At a pressure of $17.0\, $bar, the ZDW is at $803\, $nm, and at higher pressures it shifts to longer wavelengths.
Since the generated sidebands are too weak for a spectrum analyser, we implemented single-photon fiber spectroscopy to locate the frequencies of signal and idler photons.   

For the single-photon fiber spectroscopy, we used the optical setup shown in Fig. \ref{fig:setup}(a). In this method, the signal and idler photons are sent through a GVD medium~\cite{valencia2002entangled} (optical fibers 1 and 2), and then, their relative time delay in the fiber is mapped to their spectral separation. For biphotons generated by a pulsed pump, the pump pulse can serve as a time trigger. Then, different spectral components of the signal and idler photons map to different arrival times at the detectors, and the JSI can be measured~\cite{avenhaus2009fiber}.

First, we calibrated our system by measuring the travel times through optical fibers 1 and 2 for pulses with known spectra, obtained from a tunable source of bright twin beams and monitored with an OSA. A third  SNSPD registered the time of arrival of the pump photons and thus served as a reference optical trigger. Figs.~\ref{fig:setup}(c) and~\ref{fig:setup}(d) show the obtained calibration curves for the signal and idler channels, respectively. The spectral resolution of the setup, determined by the lengths of the fibers, their dispersion, and the jitter of the detectors ($120\,$ps each), was between $7$ and $11\,$THz, depending on the wavelength range.  

We then employed our calibrated experimental scheme to measure the spectral positions and the JSIs of the biphotons for different pressures of xenon.  For the JSIs, we registered  two-dimensional coincidence histograms of the arrival-time difference between the signal photons and the optical trigger, and simultaneously, the arrival-time difference between the idler photons and the optical trigger \cite{avenhaus2009fiber}. The obtained 2D-coincidence histogram in the spectral domain corresponds to the JSI of the FWM biphotons. The spectral positions were obtained with the same method but only measuring the arrival time difference between the signal and idler photons and assuming the energy conservation condition Eq. (\ref{FWMenergy}).

\begin{figure}[ht!]
\centering\includegraphics[width=0.92\linewidth]{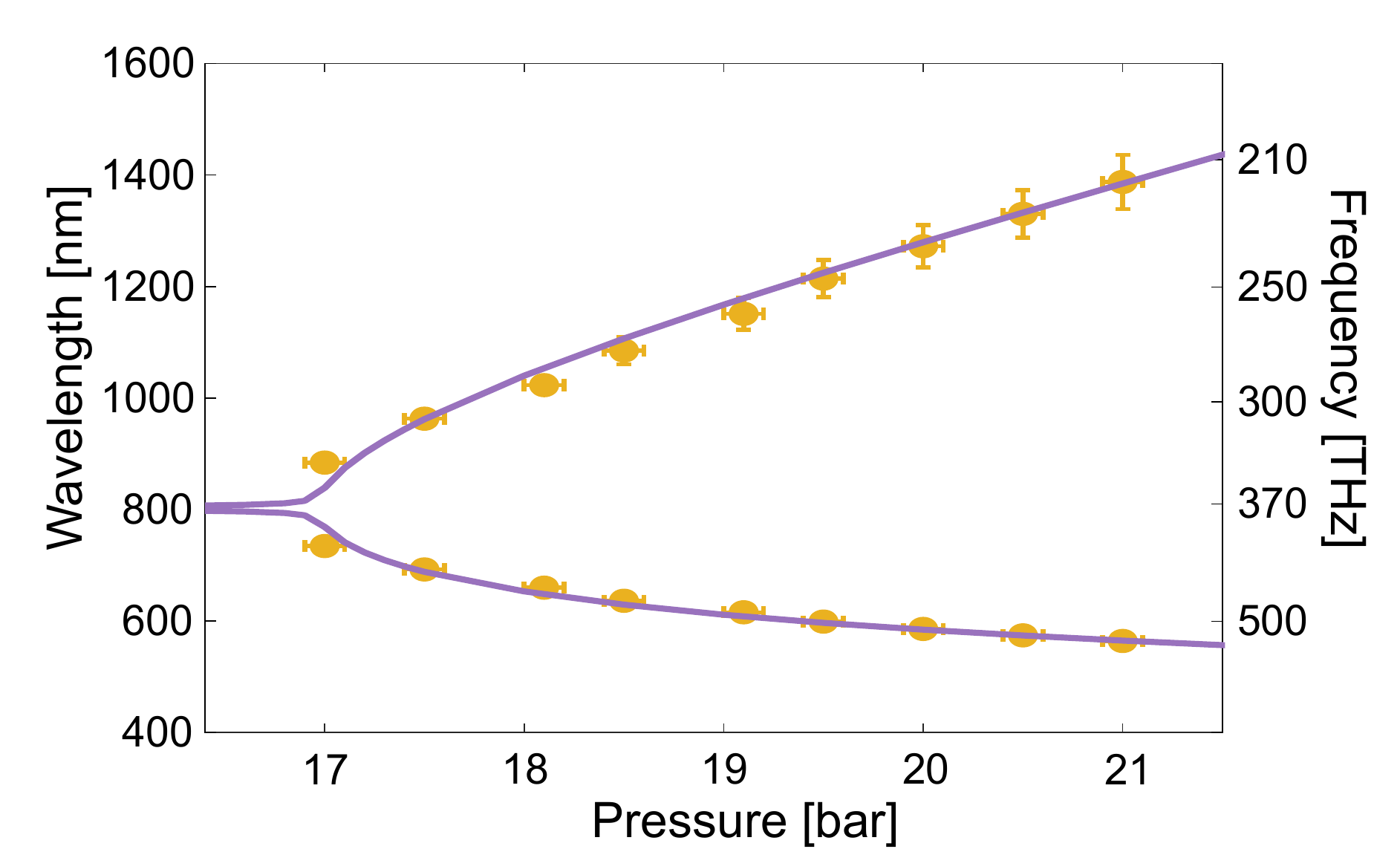}
\caption{Measured (points) and calculated (solid line) wavelengths of the signal and idler photons as functions of the gas pressure inside the kagomé PCF.}
\label{fig:tunabilitybiphotons}
\end{figure}

Fig. \ref{fig:tunabilitybiphotons} shows the obtained wavelengths (yellow points) as functions of the  gas pressure inside the  kagomé PCF.  We observe that the spectral separation between them increases along with the pressure. For a pressure range of $4\,$bar, we obtained a tunability of $30\,$THz/bar. At pressure $21.0\,$bar, the signal and idler photons are at $564$ and $1384\,$nm, respectively, more than one octave apart. This tunability range of $120\,$THz exceeds the  previously reported values~\cite{cordier2019active} by a factor of 11. The solid purple line is the result of analytical calculations based on the energy conservation and the nonlinear phase-matching conditions (Eqs.~(\ref{FWMenergy}),~(\ref{FWMmomentum})) for different pressures. The dispersion dependence of the xenon-filled kagomé was calculated using the extended version of the Marcatili-Schmeltzer model, with an s-parameter of $0.03$~\cite{Finger:14}. The experimental data and the analytical calculations agree. 

The spectral tunability of the source is limited at low pressures,  since at low powers, the signal and idler photons are spectrally very close to the $802\,\mathrm{nm}$ pump radiation (Fig. \ref{fig:tunabilitybiphotons}), making it challenging to filter them out. At higher pressures, the  efficiency of the FWM process is strongly reduced due to the larger group-velocity mismatch between the biphoton and the pump photons. Thus, at pressures above $21.0\,$bar, we only detected the intrinsic noise of the system.
\begin{figure}[ht!]
\centering\includegraphics[width=0.9\linewidth]{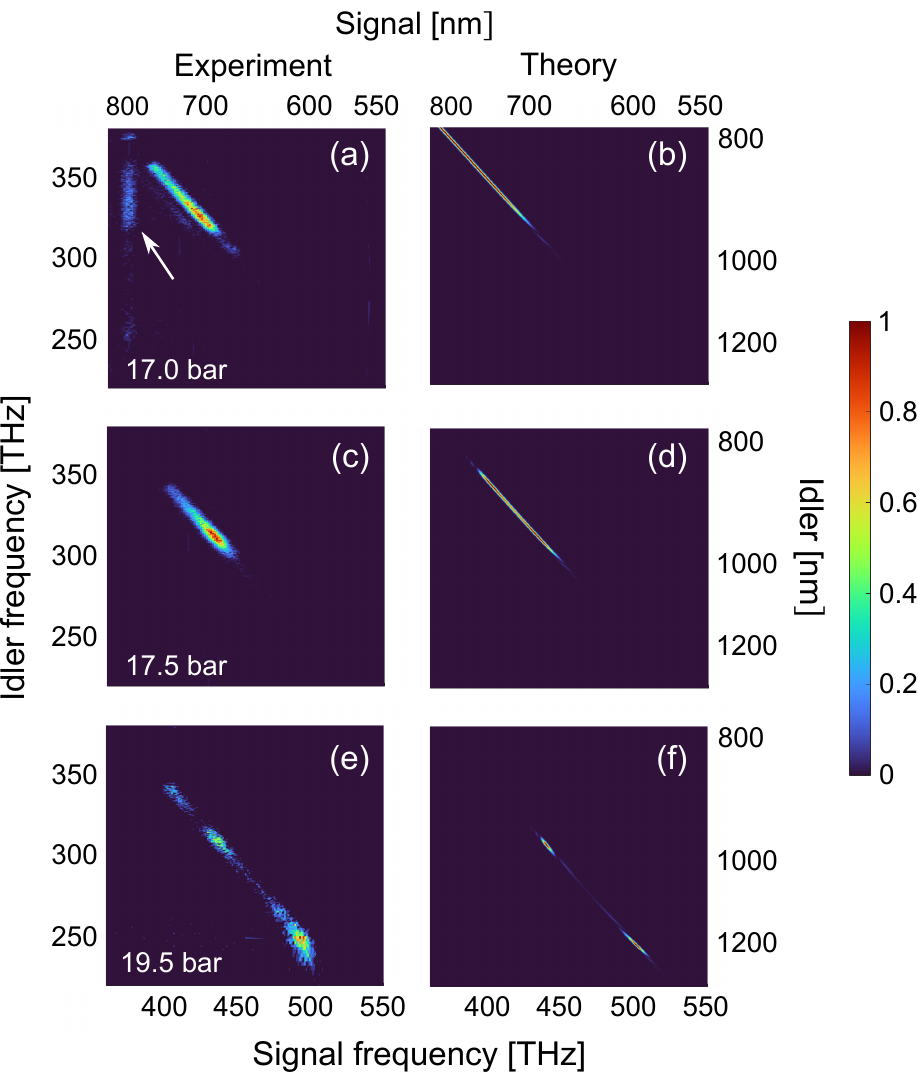}
\caption{Measured (\textit{left}) and calculated (\textit{right}) joint spectral intensities of the signal and idler photons with an average pump power of 0.2 mW for 17.0 bar (a),(b); 17.5 bar (c),(d); and 19.5 bar (e),(f). The white arrow indicates the detected unfiltered pump photons.}
\label{fig:JSI}
\end{figure}

Further, we measured the JSI and also calculated it in a similar manner as \cite{garay2008tailored}. This method allowed us to derive the JSI in a closed analytical form taking  higher-order dispersion terms of the gas-filled HC-PCF into account.
Fig.~\ref{fig:JSI} shows the experimental (left) and theoretical (right) JSIs obtained at three different pressures with an average pump power of $0.2\,\mathrm{mW}$. At $17.0\,\mathrm{bar}$ (Figs.~\ref{fig:JSI}(a, b)), one can observe a very broad shape of the JSI stretched along $-45^\circ$ with respect to the signal frequency axis. The shape of the JSI confirms spectral anti-correlations between the signal and idler photons and reveals the very broad spectral bandwidth of the biphotons.  The experimental results show good agreement with the theoretical calculations. However, the measured JSI is broadened due to the finite spectral resolution. This can be improved by using longer fibers at the cost of higher losses.  The vertical stripe centered at $374\,\mathrm{THz}$ in Fig.~\ref{fig:JSI}(a), shown by a white arrow, corresponds to the unfiltered pump photons when F1 and F2 are removed. This was done only at the pressure of $17.0$ bar to minimize cutting of the JSI by the filters. Calculation (panel b) shows that the spectrum of biphotons occupies in this case the whole spectral range around the pump.

The JSI at $17.5\,\mathrm{bar}$ is shown in Figs. \ref{fig:JSI}(c, d). Here, we can observe a reduction of the spectral bandwidth of the signal and idler photons with a full-width at half maximum (FWHM) of  $22 \pm 1\,\mathrm{THz}$ but with a spectral separation of $120\,\textrm{THz}$. A source with these characteristics can be beneficial for sensing with undetected photons since the idler wavelength, around $1000\,$nm, covers the range of inefficient detection. Besides, the large  spectral bandwidth leads to short correlation times \cite{garay2008ultrabroadband}. The experimental and theoretical results closely match, except for the resolution limitations and  the asymmetry of the experimental JSI. We attribute this asymmetry to experimental losses at short wavelengths due to the poor detection efficiency of the IR SNSPDs and the rapid increase of losses for wavelengths shorter than $800\,\mathrm{nm}$ ($2.0\,\mathrm{dB/km}$ loss at $800\,\mathrm{nm}$) in the SMF-28 fibers. This effect is also visible at $17.0\,\mathrm{bar}$.

Fig.~\ref{fig:JSI}(e) shows the JSI measured at $19.5\,\mathrm{bar}$. The JSI has an elongated shape composed of two main peaks, each of them having width more than $10$ THz and additionally broadened due to the finite resolution. The  complicated JSI shape is caused by the higher-order dispersion terms, appearing when the signal and idler photons have a very large spectral separation. Analytical calculations shown in Fig. \ref{fig:JSI}(f) reproduce this structure reasonably well. The main JSI peak is at the signal frequency $493$ THz, and it is the position of this peak that is marked in Fig. \ref{fig:tunabilitybiphotons}.

\begin{figure}[hbt!]
\centering\includegraphics[width=0.90\linewidth]{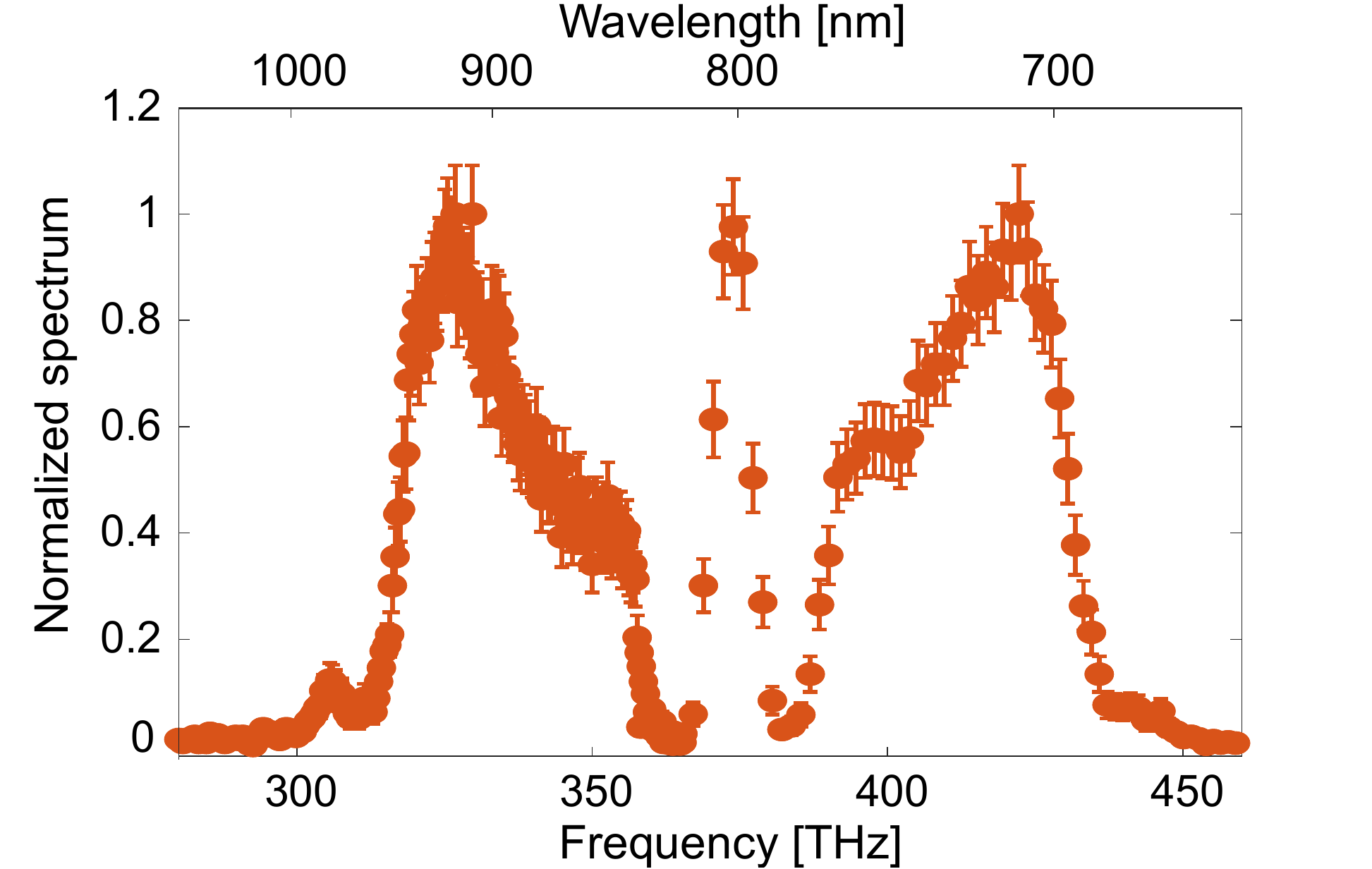}
\caption{Experimental spectrum of the signal and idler photons at a fixed pressure of $17.0\,\mathrm{bar}$.}
\label{fig:marginalspectra}
\end{figure}

The case of the lowest pressure ($17.0\,\mathrm{bar}$) deserves a special attention because (see Fig.~\ref{fig:JSI}(b)) it leads to ultra-broadband biphotons whose spectrum overlaps with the pump. Fig.~\ref{fig:marginalspectra} shows the FWM spectrum at this pressure obtained by projecting the JSI on the frequency axis. The spectrum covers a broad frequency range ($110\,$THz) around the pump, which, due to its high brightness, is seen as a peak at the center despite the notch filters. The notch filters cut part of the biphoton spectrum from $768\,$nm to $835\,$nm.

In conclusion, we have experimentally demonstrated a fiber-~based biphoton source with a high coincidence-to-accidentals ratio and an ultrabroad frequency tunability. We tuned the frequencies of signal and idler photons over $\sim120\,\mathrm{THz}$ by adjusting the pressure from $17.0$ to $21.0\,\mathrm{bar}$. To our knowledge, this is the largest tunability range ever achieved in fiber-based biphoton sources. Both the frequencies of signal and idler photons and their spectral bandwidths are tuned without changing the pump frequency or the fiber, but by simply varying the gas pressure inside the  HC-PCF~\cite{finger2017characterization}. All obtained spectra are very broad, covering more than $20\,$THz and in some cases manifesting complicated structures. This source will be useful for OCT and other types of sensing with undetected photons. 

\smallskip \textbf{Funding} We acknowledge the financial support by Deutsche Forschungsgemeinschaft (DFG) (Grants No. CH-1591/9-1 and No. JO-1090/6-1). \\
\textbf{Disclosures} The authors declare no conflicts of interest.


\bibliography{sample}

\end{document}